\documentclass[cameraready]{Interspeech}
\usepackage{algorithm}        
\usepackage{algorithmic} 
\usepackage{multirow}
\usepackage{multicol}
\usepackage{makecell}

\title{Zero-Shot Parkinson's Disease Detection from Speech: Comparing Large Audio and Language Models}

\author[affiliation={1}, orcid=0000-0002-6798-6535, correspondingauthor]{Muhammad Ashad}{Kabir}
\author[affiliation={2}]{Sirajam}{Munira}

\address{
    $^1$ School of Computing, Mathematics and Engineering, Charles Sturt University, NSW, Australia\\
    $^2$ Department of Computer Science, Rensselaer Polytechnic Institute, NY, USA
}

\email{akabir@csu.edu.au, munirs@rpi.edu}

\keywords{Parkinson's disease; zero-shot; large language models; large audio language models; multilingual; paralinguistics; clinical speech processing; foundation models}

\usepackage{comment}

\begin{document}

\maketitle

\begin{abstract}
Large audio and language models have recently demonstrated zero-shot reasoning capabilities across various domains. However, it remains unclear how the form of audio input, whether handcrafted acoustic features extracted from speech or the raw audio waveform itself, affects performance for Parkinson's disease (PD) detection across different languages. In this study, we systematically compare two input modalities for zero-shot PD detection: (i) handcrafted acoustic features extracted from speech recordings analyzed by a general-purpose LLM, and (ii) direct waveform input analyzed by audio-capable models. Experiments on PD speech datasets in four languages show that performance varies across input modalities, speech tasks, and languages. Handcrafted acoustic features provide more stable performance in a low-resource language (e.g., Bengali), whereas audio input yields dataset-dependent gains. These findings highlight the impact of input modality on zero-shot PD detection from speech.
\end{abstract}

\section{Introduction}

Parkinson's disease (PD) is a progressive neurodegenerative disorder characterized by both motor and non-motor impairments, including bradykinesia, rigidity, tremor, cognitive decline, mood disorders, and autonomic dysfunction~\cite{bloem2021parkinson}. Globally, PD affects more than 10 million people~\cite{parkinsonStatistics} and represents the fastest-growing neurological disorder in terms of prevalence and disability burden~\cite{steinmetz2024global}. Early detection remains critical, as timely therapeutic intervention can significantly improve symptom management, slow functional deterioration, and enhance quality of life, particularly during the prodromal and early stages~\cite{bloem2021parkinson, postuma2015mds}. Among the early and consistently reported signs of PD are abnormalities in speech production, collectively referred to as hypokinetic dysarthria~\cite{ramig2008speech}, which affect up to 90\% of individuals~\cite{ho1999speech} over the disease period. These speech impairments include reduced vocal intensity (hypophonia), monotonic prosody, articulatory imprecision, increased jitter and shimmer, altered harmonic-to-noise ratio, and disrupted temporal speech patterns~\cite{ramig2008speech, little2008suitability, skodda2008speech}. Because speech can be collected non-invasively, remotely, and at low cost using widely available recording devices, it has emerged as a promising digital biomarker for scalable and continuous PD assessment~\cite{tsanas2012novel}. 

Over the past decade, speech-based PD detection has emerged as one of the most actively studied digital biomarker paradigms, as documented in multiple systematic and narrative reviews of computational speech analysis in PD~\cite{mei2021machine, tabashum2024machine}. Foundational studies demonstrated that acoustic perturbation and spectral biomarkers, such as jitter, MFCCs, and prosodic descriptors, enable reliable discrimination between PD and healthy controls~\cite{little2008suitability, tsanas2012novel, sakar2019comparative, orozco2016automatic}. Existing studies established a structured methodological pipeline consisting of feature extraction, statistical feature selection, and supervised classification using machine learning (ML) models, setting the dominant paradigm for speech-based PD detection. Later studies moved beyond feature pipelines by applying deep neural networks directly to speech signals, including convolutional and recurrent architectures for phonation and connected speech analysis~\cite{farago2023cnn, vasquez2018multitask, vasquez2018multimodal}. Across both classical ML and deep learning (DL) approaches, performance improvements have largely depended on careful dataset-specific optimization, feature engineering, hyperparameter tuning, and leakage-safe validation strategies. Recent extensions towards spontaneous speech and multilingual analysis~\cite{vasquez2018towards, rusz2011quantitative, hlavnivcka2017automated, orozco2014new, hossain2025bensparx} have demonstrated that conversational speech biomarkers can reliably capture Parkinson impairments beyond sustained phonation tasks. These studies show that comprehensive acoustic feature integration combined with structured modeling strategies can achieve strong classification performance within supervised frameworks.

\begin{figure*}[!ht]
    \centering
    \includegraphics[width=.87\linewidth]{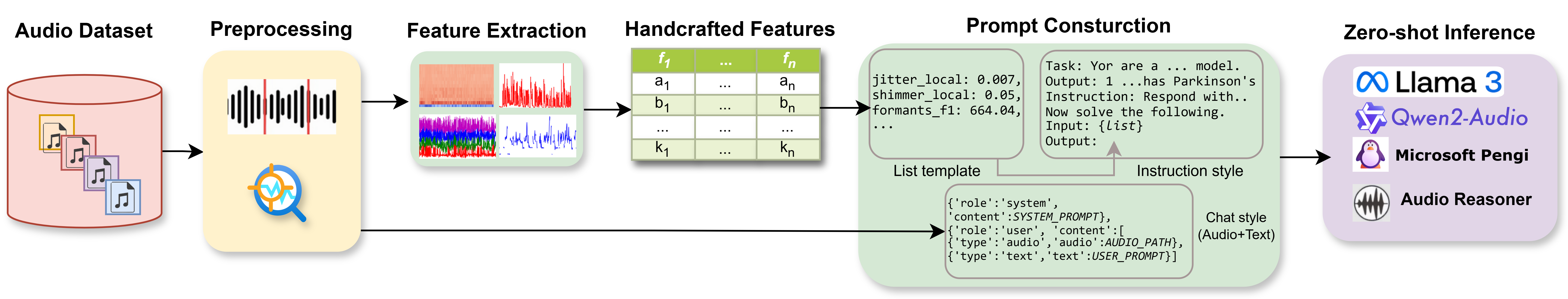}
    \caption{Schematic overview of the zero-shot pipeline for PD detection using LLMs and LALMs}
    \label{fig:methodology}
\end{figure*}
With the rapid emergence of large language models (LLMs), PD research has begun exploring their role across multiple clinical and computational paradigms. Several works employ prompting and fine-tuning strategies on structured datasets to perform diagnosis, progression modeling, or severity estimation using clinical variables~\cite{shin2026prompting, gao2024llms}. LLMs have also been used to analyze patient speech transcripts to detect medication states and neuropsychiatric changes~\cite{castelli2025detecting}. In addition, LLMs are being applied to Parkinson's care systems~\cite{zhang2025leveraging}, supporting personalized treatment planning and wearable-integrated monitoring~\cite{bhalala2025parka, cardenas2024autohealth}, as well as the construction of structured PD knowledge bases~\cite{bouchouras2024llms}. More recently, large audio language models (LALMs) have demonstrated zero-shot reasoning capabilities directly over raw speech signals, highlighting the broader feasibility of prompt-based inference from speech~\cite{shahin2025zero, peng2025survey}. 

In speech-based computational modeling, the representation of speech plays a central role in how models process clinically relevant signals. Traditionally, speech representations have been categorized into (i) handcrafted acoustic features set~\cite{little2008suitability, tsanas2012novel, eyben2010opensmile}, which provide structured and interpretable descriptors of vocal perturbations, spectral properties, and prosodic patterns, and (ii) learned representations~\cite{hinton2012deep}, such as deep neural network embeddings or raw waveform inputs~\cite{baevski2020wav2vec} that allow models to extract task-relevant features internally. With the emergence of LLM-based systems, these representation choices lead to two practical approaches: providing structured acoustic features as textual input to general-purpose LLMs or directly providing raw audio to LALMs.

Despite these developments, it remains unclear how such input choices influence zero-shot LLM performance in PD detection. Most existing PD detection research relies on supervised machine learning, whereas LLM-based studies have largely been applied to care systems and clinical support rather than direct diagnostic inference from speech. To date, there has been no systematic investigation comparing zero-shot LLM inference across different speech representations, namely structured acoustic features versus raw waveform inputs, within a unified experimental framework across multiple languages.

In this study, we evaluate the zero-shot diagnostic capability of LLMs under two distinct input paradigms: (i) extracted acoustic features from speech as input to a general-purpose LLM, (ii) direct waveform input processed through a dedicated LALM architecture. Rather than solely benchmarking performance against supervised machine learning models, our objective is to analyze how input modality influences LLM reasoning behavior, robustness, and multi-dataset generalization. Through this comparative investigation, we aim to understand how LLM performance changes depending on whether the input is raw audio or extracted acoustic features. This analysis helps clarify the strengths, limitations, and practical role of LLMs in speech-based PD detection.

\section{Methodology}
Figure~\ref{fig:methodology} provides an overview of the proposed zero-shot pipeline for PD Screening using LLMs and LALMs. The workflow consists of four main steps: (i) dataset preprocessing, (ii) extracting handcrafted features, (iii) prompt construction, and (iv) zero-shot inference.

\subsection{Datasets}

To investigate how input modality influences the zero-shot inference capability of LLMs and LALMs, we conducted experiments on four PD speech datasets, summarized in Table~\ref{tab:dataset}. The selected datasets vary in linguistic background, recording paradigm, and task constraints, enabling evaluation of modality robustness across heterogeneous conditions.
\begin{table}[!b]
    \centering
    \caption{Summary of the datasets used in this study. PD: Parkinson's disease; HC: healthy control.}
    \label{tab:dataset}
     \setlength{\tabcolsep}{3pt}
    \begin{tabular}{llrrl}
    \hline
    \multirow{2}{*}{\textbf{Dataset}} & \multirow{2}{*}{\textbf{Language}} & \multicolumn{2}{c}{\textbf{Count}} & \multirow{2}{*}{\textbf{Task}}\\
    \cline{3-4}
    & & \textbf{PD} & \textbf{HC} & \\
    \hline
    BenSParX~\cite{hossain2025bensparx} & Bengali & 60 & 60 & Conversation\\
    MDVR-KCL~\cite{jaeger2019mobile} & English & 16 & 21 & Reading text\\
    IPVS~\cite{dimauro2017assessment} & Italian & 28 & 22 & Text dependent utterance\\
    NeuroVoz~\cite{mendes2024neurovoz} & Spanish & 23 & 53 & Spontaneous speech\\

    \bottomrule
    \end{tabular}
\end{table}

The BenSParX~\cite{hossain2025bensparx} dataset comprises 60 individuals diagnosed with PD and 60 healthy control (HC) speakers, all native Bengali speakers, performing a conversational speech task. The MDVR-KCL~\cite{jaeger2019mobile} dataset includes 16 PD patients and 21 HC producing read speech under controlled conditions. The IPVS~\cite{dimauro2017assessment} dataset contains recordings from 28 PD patients and 22 HC performing text-dependent utterances, where predefined sentences were repeated. NeuroVoz~\cite{mendes2024neurovoz} dataset includes 23 PD patients and 53 HC producing spontaneous (free) speech.

\subsection{Data Preprocessing and Feature Extraction}
All speech recordings were processed using a standardized pipeline to ensure acoustic consistency across datasets. Background noise was attenuated while preserving salient vocal characteristics, and all recordings were resampled to 16 kHz to eliminate sampling-rate variability. 

\begin{table*}[!ht]
    \centering
    \caption{Prompt templates used for zero-shot Parkinson's disease classification. \{list\} denotes the serialized handcrafted acoustic feature-value pairs (e.g., jitter\_local: 0.007).}
    \label{tab:prompt}
    \begin{tabular}{lp{15cm}}
     \toprule
      \textbf{Model type}   & \textbf{Prompt template} \\
      \midrule
      LLM   & Task: You are a clinical classification model. Based on the audio features extracted from a person's speech, classify whether a person has Parkinson's disease or not. Output 1 if the person has Parkinson's disease, or 0 if the person is healthy.
Instruction: Respond with exactly one token: 0 or 1.
Now solve the following.
Input: \{\textit{list}\}
Output:\\
\hline
\makecell[t l]{LALM,\\LARM} & You are an audio analysis model. Your task is to decide whether the speech characteristics are more consistent with healthy control speech or Parkinson's. Consider acoustic cues, including pitch variability, loudness variability over time, articulation precision of consonants, voice quality (breathy, hoarse, strained), speech rate, and rhythm. Make a balanced decision based only on the provided audio. 
Output only a single digit: 0 = Healthy or 1 = Parkinson's disease.\\
     \bottomrule
    \end{tabular}
\end{table*}

For extracting handcrafted acoustic features, each recording was segmented into non-overlapping 10-second intervals, from which 71 handcrafted acoustic features were extracted following~\cite{hossain2025bensparx}. The feature set includes time-domain, frequency-domain, and cepstral descriptors associated with Parkinsonian speech, such as jitter, shimmer, harmonic-to-noise ratio, and Mel-frequency cepstral coefficients. An identical feature extraction configuration was used across all corpora without dataset-specific modification, ensuring that performance differences reflect input modality rather than variations in features.
\subsection{Large Audio and Language Models}
To evaluate the effect of input modality on zero-shot inference, we employed three categories of large-scale models without task-specific fine-tuning: LLMs, LALMs, and large audio reasoning models (LARMs) (a reasoning-enhanced subset of LALMs).
For feature-based inference, we used LLaMA 3 (8B)~\cite{meta2024llama3}, a decoder-only transformer LLM for text generation. The 71-dimensional handcrafted acoustic features were serialized into structured prompts and provided to the model in a zero-shot setting.

For direct audio inference, we used Qwen2-Audio~\cite{chu2024qwen2}, a LALM that follows a speech-and-text-to-text paradigm. It combines an audio encoder (e.g., Whisper-large-v3) with a Qwen LLM backbone to process waveform inputs alongside textual instructions. Raw audio signals were provided with a prompt for zero-shot PD inference.
We also evaluated Pengi~\cite{deshmukh2023pengi}, a LALM that represents audio as continuous embeddings from a pretrained encoder and integrates them with textual prompts as input to a frozen LLM, enabling zero-shot prediction from raw waveform recordings.

Finally, we included Audio-Reasoner~\cite{xie2025audio}, a LARM designed for higher-level reasoning over acoustic signals via cross-modal alignment between audio encoders and LLMs. Audio inputs were directly provided with a binary prompt for zero-shot inference.

\subsection{Prompt Construction}
In the prompt construction stage, handcrafted acoustic features are serialized into a structured textual \textit{list} of feature–value pairs (see Figure~\ref{fig:methodology}), converting tabular feature vectors into a format compatible with LLMs. The serialized \textit{list} is embedded within a standardized instruction-style prompt specifying the task, output constraints, and response format (see Table~\ref{tab:prompt}). For text-based LLMs, no additional task-specific clinical context beyond the provided features is included.

For audio-capable LLMs (e.g., Qwen2-Audio, Pengi, and Audio-Reasoner), a multimodal (audio + text) prompt template is used, where the same task instruction (Table~\ref{tab:prompt}) is paired with raw waveform input in a role-based (system/user) format (Figure~\ref{fig:methodology}). This prompt includes general guidance on clinically relevant acoustic cues (e.g., pitch variability, and voice quality). 
The prompt text was designed based on standard clinical descriptions of speech characteristics associated with Parkinson’s disease and iteratively refined for clarity and consistency.

\section{Experiments and Results}

\subsection{Experiments}

We evaluated four large-language and audio-language models under a unified zero-shot framework. LLaMA 3 (8B)\footnote{{\scriptsize\url{https://huggingface.co/meta-llama/Meta-Llama-3-8B}}} and Qwen2-Audio (7B-Instruct)\footnote{\scriptsize\url{https://huggingface.co/Qwen/Qwen2-Audio-7B-Instruct}} were obtained from the Hugging Face repository. Pengi\footnote{\scriptsize\url{https://github.com/microsoft/Pengi/tree/main}} and Audio-Reasoner\footnote{\scriptsize\url{https://github.com/xzf-thu/Audio-Reasoner}} were implemented from their official repositories with default inference configurations. 
All experiments used a fixed random seed (0) and deterministic decoding (temperature = 0), with each recording evaluated once per model.
Predictions for LLaMA 3 were derived from token-level log probabilities of candidate labels, whereas Pengi, Audio-Reasoner, and Qwen2-Audio produced explicit generated predictions with associated probabilities. 

For the handcrafted acoustic feature set using text-based LLMs (e.g., LLaMA 3), segment-level predictions were aggregated to obtain subject-level decisions in order to ensure clinically meaningful evaluation and a consistent diagnostic scale across modalities. For each subject, the final label was determined by majority vote across segment predictions. In the case of a tie, the label with the higher mean predicted probability across the corresponding segments was selected. The subject-level probability was computed as the mean probability of segments assigned to the final predicted label.

Model performance was assessed at the subject level using balanced accuracy, AUROC, sensitivity, specificity, and Brier score. To quantify statistical uncertainty, 95\% confidence intervals were estimated using stratified bias-corrected and accelerated nonparametric bootstrap resampling with 10,000 replicates, preserving class proportions in each bootstrap sample. Experiments were conducted on a machine equipped with an NVIDIA GeForce RTX 3080 GPU and 16 GB of memory.

\subsection{Results}

\begin{table*}[!ht]
    \centering
    \caption{Zero-shot subject-level performance across input modalities (handcrafted acoustic features and raw waveform) for PD detection, reported as balanced accuracy, AUROC, sensitivity, specificity, and Brier score with 95\% bootstrap confidence intervals.}
    \label{tab:result}
    \setlength{\tabcolsep}{6pt}
    \begin{tabular}{l ll ccccc}
    \toprule
     \textbf{Dataset} & \multicolumn{2}{c}{\textbf{Model}} & \textbf{B. Accuracy} (\%)  & \textbf{AUROC}  & \textbf{Sensitivity} (\%) & \textbf{Specificity} (\%)  & \textbf{Brier Score}   \\ \cmidrule{2-3}
     & \textbf{Type} & \textbf{Name} & ($\uparrow$) & ($\uparrow$) & ($\uparrow$)  & ($\uparrow$) & ($\downarrow$)\\
     \midrule
    BenSParX 
          & LLM &  LLaMA 3 & $83.33_{\text{75.83--89.17}}$ & $0.901_{\text{0.827--0.947}}$ & $86.67_{\text{73.33--91.67}}$ & $80.00_{\text{66.67--88.33}}$ & $0.228_{\text{0.225--0.233}}$\\
          & LALM &  Qwen2-Audio & $50.00_{\text{42.50--55.83}}$ & $0.536_{\text{0.429--0.638}}$ & $16.67_{\text{6.67--25.00}}$ & $83.33_{\text{70.00--90.00}}$ & $0.258_{\text{0.246--0.272}}$ \\
          & LALM &  Pengi & $58.33_{\text{49.17--66.67}}$ & $0.617_{\text{0.513--0.712}}$ & $45.00_{\text{31.67--56.67}}$ & $71.67_{\text{58.33--80.00}}$ & $0.258_{\text{0.225--0.296}}$ \\
          & LARM  & Audio-Reasoner & $50.83_{\text{45.00--55.00}}$ & $0.534_{\text{0.436--0.628}}$ & $91.67_{\text{80.00-96.67}}$ & $10.00_{\text{3.33--18.33}}$ & $0.396_{\text{0.352--0.444}}$ \\
            \hline
    MDVR-KCL 
          & LLM &  LLaMA 3 & $50.74_{\text{42.86--59.38}}$ & $0.702_{\text{0.503--0.848}}$ & $6.25_{\text{0.00-18.75}}$ & $95.24_{\text{71.43--100.00}}$ & $0.242_{\text{0.236--0.247}}$ \\
          & LALM &  Qwen2-Audio & $52.98_{\text{39.58--66.37}}$ & $0.506_{\text{0.308--0.696}}$ & $25.00_{\text{6.25--43.75}}$ & $80.95_{\text{52.38--90.48}}$ & $0.248_{\text{0.229--0.266}}$ \\
          & LALM &  Pengi & $29.76_{\text{14.88--42.41}}$ & $0.268_{\text{0.122--0.473}}$ & $50.00_{\text{18.75--68.75}}$ & $9.52_{\text{0.00--23.81}}$ & $0.442_{\text{0.364--0.521}}$ \\
          & LARM & Audio-Reasoner & $69.49_{\text{56.12--81.25}}$ & $0.609_{\text{0.414--0.783}}$ & $43.75_{\text{12.50--62.50}}$ & $95.24_{\text{71.43--100.00}}$ & $0.252_{\text{0.159--0.354}}$ \\
            \hline 
    IPVS 
          & LLM &  LLaMA 3 & $51.79_{\text{50.00--55.36}}$ & $0.805_{\text{0.659--0.904}}$ & $3.57_{\text{0.00-10.71}}$ & $100.0_{\text{100.0--100.0}}$ & $0.249_{\text{0.246--0.253}}$ \\
          & LALM &  Qwen2-Audio & $54.87_{\text{40.42--67.86}}$ & $0.472_{\text{0.305--0.634}}$ & $64.29_{\text{39.29--78.57}}$ & $45.45_{\text{22.73--63.64}}$ & $0.254_{\text{0.238--0.270}}$ \\
          & LALM &  Pengi & $32.63_{\text{19.97--43.99}}$ & $0.188_{\text{0.083--0.341}}$ & $10.71_{\text{0.00-21.43}}$ & $54.55_{\text{27.27--68.18}}$ & $0.444_{\text{0.381--0.507}}$\\
          & LARM & Audio-Reasoner & $51.95_{\text{42.05--62.34}}$ & $0.395_{\text{0.261-0.542}}$ & $85.71_{\text{64.29--92.86}}$ & $18.18_{\text{4.55--31.82}}$ & $0.398_{\text{0.307--0.498}}$ \\
            \hline 
    NeuroVoz 
         &  LLM &  LLaMA 3 & $52.58_{\text{41.06--64.68}}$ & $0.486_{\text{0.342--0.635}}$ & $39.13_{\text{17.39--56.52}}$ & $66.04_{\text{50.94--75.47}}$ & $0.247_{\text{0.240--0.254}}$ \\
         &  LALM &  Qwen2-Audio & $63.04_{\text{54.35--71.74}}$ & $0.519_{\text{0.355--0.685}}$ & $26.09_{\text{8.69--43.48}}$ & $100.0_{\text{100.0--100.0}}$ & $0.219_{\text{0.205--0.235}}$ \\
         &  LALM &  Pengi & $53.04_{\text{45.28--63.14}}$ & $0.517_{\text{0.378--0.656}}$ & $17.39_{\text{4.35--30.43}}$ & $88.68_{\text{73.58--94.34}}$ & $0.235_{\text{0.201--0.271}}$ \\
         &  LARM & Audio-Reasoner & $63.90_{\text{53.32--75.39}}$ & $0.676_{\text{0.525--0.797}}$ & $39.13_{\text{17.39--56.52}}$ & $88.68_{\text{75.47--94.34}}$ & $0.207_{\text{0.152--0.270}}$ \\
\bottomrule
\multicolumn{8}{l}{B. Accuracy: Balanced Accuracy}
    \end{tabular}

\end{table*}

Table~\ref{tab:result} summarizes the subject-level performance across datasets and model categories, enabling a comparison between handcrafted acoustic features analyzed by a LLM and raw waveform input processed by LALM and LARM models for zero-shot PD inference.

For the BenSParX dataset, handcrafted features-based inference using LLaMA 3 achieved the highest observed balanced accuracy (83.3\%) and AUROC (0.901), along with a lower Brier score (0.228), indicating both strong discrimination and relatively well-calibrated predictions. In contrast, audio-based models yielded near chance-level balanced accuracy (50--58\%) and exhibited weaker discrimination (AUROC 0.534--0.617). Audio-Reasoner showed high sensitivity (91.67\%) but very low specificity (10\%), indicating a strong bias toward positive predictions and correspondingly high Brier loss (0.396). Overall, feature-based input provided more stable performance on this low-resource language dataset (i.e., Bengali), although confidence intervals overlap and differences should be interpreted with caution.

On MDVR-KCL and NeuroVoz, however, the gap between modalities narrowed. For MDVR-KCL, performance patterns were more heterogeneous. Audio-Reasoner achieved the highest balanced accuracy (69.49\%), exceeding LLaMA 3 and other audio models. However, LLaMA 3 maintained the highest AUROC (0.702) and comparable calibration (Brier 0.242), indicating better overall ranking performance despite modest balanced accuracy. Pengi showed the weakest overall discrimination on this dataset (balanced accuracy 29.76\%, AUROC 0.268) with the highest Brier score. These results suggest that under controlled read speech conditions, certain audio-based reasoning models may benefit from raw waveform information, although discrimination consistency remains variable. Similarly, for the NeuroVoz dataset, audio-based models showed improved relative performance. Audio-Reasoner and Qwen2-Audio achieved the highest balanced accuracies (63.9\% and 63.04\%, respectively), outperforming LLaMA 3. Audio-Reasoner also achieved the highest AUROC (0.676) and the lowest Brier score (0.207), indicating comparatively better calibrated predictions in this dataset. Here, direct audio input appeared more advantageous than handcrafted features.

In contrast, for the IPVS dataset, overall performance was moderate across models. Qwen2-Audio achieved the highest balanced accuracy (54.87\%), while LLaMA 3 reached the highest AUROC (0.805). Pengi again produced the weakest results with low balanced accuracy (32.63\%) and AUROC (0.188), indicating weak zero-shot generalization in this setting. Notably, specificity was extremely high for LLaMA 3 (100\%) but accompanied by very low sensitivity (3.57\%), reflecting a strong bias toward the majority class. 

Overall, the results suggest that input modality influences zero-shot PD detection performance, although differences vary across datasets. Handcrafted acoustic features yielded more consistent performance across datasets, with stronger discrimination and lower Brier scores (indicating better-calibrated predictions) in some cases, including the Bengali dataset. In contrast, raw waveform input improved balanced accuracy in some datasets but showed greater variability across settings. Rather than demonstrating universal superiority of one modality, these findings indicate that feature-based prompting may provide more stable zero-shot behavior, while waveform-based input yields dataset-dependent results. These findings may also be influenced by factors such as model architecture, dataset characteristics, and recording conditions, and should therefore be interpreted with caution.

\section{Discussion}
The results from this study suggest that input modalities (direct audio waveform vs. handcrafted acoustic features) influence how zero-shot LLM systems process and interpret speech-based clinical signals. Rather than reflecting a uniform performance hierarchy, the observed patterns indicate that different input modalities interact differently with model architectures. Handcrafted acoustic features provide explicit summaries of clinically relevant vocal characteristics, which may better align with the reasoning structure of text-based LLMs. In contrast, audio-capable models must implicitly extract task-relevant cues from raw waveforms, potentially introducing variability when operating without task-specific adaptation. Importantly, improvements in balanced accuracy under raw audio input were not consistently accompanied by lower Brier scores, indicating that higher discrimination does not always correspond to better-calibrated predictions. This suggests that zero-shot inference from waveform data may capture useful signals, yet still produce less stable probability estimates. 
As prior work has noted, the use of LLMs in clinical decision-making raises ethical concerns regarding reliability, bias, and transparency~\cite{li2023ethics,freyer2024future}. Therefore, these models should not be used for clinical diagnosis without rigorous prospective validation~\cite{de2024text} and appropriate regulatory and clinical oversight~\cite{ong2024ethical}.


Several limitations should be acknowledged. First, all evaluations were conducted in a strictly zero-shot setting without task-specific fine-tuning or in-context (few-shot) learning. While this design isolates modality effects, it may underestimate the potential of audio-capable models when adapted to speech data. Second, dataset sizes remain modest, particularly for certain languages, which may affect the stability of performance estimates despite bootstrap confidence intervals. Finally, handcrafted features were derived from a predefined pipeline rather than optimized individually for each corpus, which may limit modality-specific adaptation.

\section{Conclusion}
This study examines the impacts of zero-shot speech-based PD detection, comparing handcrafted acoustic features analyzed by a text-based LLM with raw waveform input processed by LALMs. Across four datasets in four different languages under a unified evaluation protocol, we observed that model performance is modality-dependent: feature-based prompting yielded the most stable results on the Bengali corpus, whereas direct audio input produced dataset-specific gains but with higher variability and less consistent calibration. These findings reinforce that zero-shot capability alone is insufficient to ensure robust PD screening, and future work should explore in-context learning, fine-tuning, reliability, and prospective validation.

\section{Generative AI Use Disclosure}

ChatGPT (version 5.2, OpenAI) was used for language editing and refinement of the manuscript. 

\bibliographystyle{IEEEtran}
\bibliography{mybib}

\end{document}